# An *Ab Initio* Cluster Study of Atomic Oxygen Chemisorption on Ga-Rich GaAs(100) (2 × 1) and *β*(4 × 2) Surfaces

Michael L. Mayo and Asok K. Ray*

*Department of Physics, University of Texas at Arlington, Arlington, Texas 76019*

## Abstract

*Ab initio* self-consistent total energy calculations using second order Møller-Plesset perturbation theory and Hay-Wadt effective core potentials for gallium and arsenic have been used to investigate the chemisorption of atomic oxygen on the Ga-rich GaAs (100) (2 × 1) and *β*(4 × 2) surfaces. Finite sized hydrogen saturated clusters with the experimental zinc-blende lattice constant of 5.654Å and the energy optimized surface Ga dimer bond length of 2.758Å have been used to model the semiconductor surface. We present the energetics of chemisorption on the (100) surface layer including adsorption beneath the surface layer at two interstitial sites. Chemisorption energies, nearest surface neighbor Ga-O bond lengths, and homo-lumo gaps are reported for all considered sites of chemisorption and compared with published results in the literature on O adsorption on the GaAs surface. Results are also compared with our previous results on hydrogen chemisorption on the same GaAs surface. Possibilities of transition of the surface from a semi-conducting state to a semi-insulating state are also discussed.

## I. Introduction

The technological applications of GaAs due to its high electron mobility and direct band gap make it an important system for fundamental and applied research. The industry standard for growing GaAs by molecular beam epitaxy (MBE) is the (100) surface. This surface has the highest aerial density of dangling surface bonds, greater than the (110) or (111) surfaces and consequently, surface reconstruction is facilitated by these bonds. In this work, we extend our previous work on cesium and hydrogen chemisorption on the Ga-rich GaAs(100) surface [1] to study oxygen chemisorption on such a surface. As is known, oxygen adsorption on the GaAs semiconductor surface has been the subject of ongoing research and great debate for several decades due to the importance of GaAs in the semiconductor industry as well as the high contamination rate of oxygen on the GaAs surface. This is partly due to the chemical character of oxygen, the predominant physical trait responsible for the controversy over the correct characterization of this system. For the sake of brevity, we first briefly comment only on *some* of the published literature [2-26].

*email:akr@uta.edu

Ismail et al. [5] studied the electronic properties of cleaved GaAs(110) surface induced by oxygen adsorption by contact potential difference measurement. The oxygen exposure induced acceptor and donor surface states pinning the surface Fermi level at approximately 0.45 and 0.7eV above the valence band for n- and p-doped samples, respectively. Noticeable modifications of the electron affinity were also produced. Stroscio et al. [6] reported, from scanning tunneling microscopy results, on the initial phase of oxygen adsorption on p-type GaAs(110) surface. Atomic oxygen was found to be bonded in an inter-chain bridging position roughly equidistant from a single As atom and two Ga Atoms in a neighboring chain. No evidence for band bending was found and was in marked contrast to previous results on n-type GaAs(110) surface. Atomic force microscopy was used by Nayak et al. [15] to investigate the influence of controlled oxygen incorporation on the surface morphology of GaAs films grown by metallorganic vapor phase epitaxy. Oxygen was found to influence the periodic morphology leading to a breakup and a model was proposed in which oxygen preferentially attached at steps. Lee et al. [16] investigated, by x-ray photoelectron diffraction (XPD) analysis, the initial stages of oxidation on the GaAs(110) surface. They concluded that the initial oxidation occurred only at the surface, and that the subsurface oxidation does not take place. The angular variation of the intensity ratio between the oxide and the bulk components of the As core levels showed one specific bonding unit corresponding to the double bonding of oxygen atoms with As atoms in the direction of 55-60 degrees relative to the surface normal in the twofold symmetry plane. Mattila and Nieminen [17] studied oxygen point defects with the plane-wave pseudopotential method in GaAs, GaN, and AlN. As they point out, theoretical modeling of oxygen in semiconductor materials is a computational challenge due to the chemical character of oxygen, such as bonding properties, large electronegativity, and sharp character of the oxygen electronic wave function. Taguchi and Kageshima [21] investigated the oxygen negative-$U$ center in GaAs using the local-density-functional approximation of density functional theory on two different As-rich GaAs surfaces. Five charge states of oxygen from +1 to -3 were considered in the O-$V_{As}$ and $O_I$ structures. For the O-$V_{As}$ structure, as the Fermi level rises, the state having the lowest formation energy was found to change from 0, -1, -2, and -3. Hence, the O-$V_{As}$ structure may take four charge states from 0 to -3. Four charge states from 0, -1, -2, and -3 were investigated for the $O_I$ structure. The +1 charge state was found to be unstable as the O atom moves toward the center Ga-$O_i$-Ga site. Both sites were found to favor Ga-O-Ga bonding. Accurate total-energy pseudopotential methods have been used by Pesola et al. [22] to study the structures, binding energies, and the local vibrational modes of various models for the Ga-O-Ga defect in GaAs. In contrast with previous models, a new model $(As_{Ga})_2$-$O_{As}$ was introduced, the properties of which were found to be in agreement with experimental observations. Orellana and Ferraz [24] investigated the structural properties, formation energies, and electronic structure of oxygen impurities in GaAs using first-principles total-energy calculations. For the substitutional defect, they found stable positions for the O atom with $C_{2v}$ symmetry for the 2- and the 3- charge states. For the 1- charge state, they found two equilibrium positions, with $T_d$ and $C_{2v}$ symmetries, which were very close in energy. For the interstitial defect also, they found three equilibrium positions for the O atom. S. I. Yi et al. [26] examined adsorption of atomic oxygen on As-rich GaAs(001)-(2 x 4) and the resulting surface structures. Experiments were carried out using an ultrahigh vacuum chamber equipped with a low energy electron diffraction spectrometer, an Auger electron spectrometer, a mass spectrometer and a scanning tunneling microscope capable of imaging at atomic resolution. The GaAs (001)-(2 x 4) surface was prepared by molecular beam epitaxy. They observed As atoms dislodging from the top layer As dimers at the onset of oxidation. The structure of an oxygen occupied site was studied using density functional theory to perform geometry optimization on $Ga_{20}As_{20}H_{32}$, $Ga_{20}As_{19}H_{32}+O$, and $Ga_{20}As_{19}H_{32}+O^-$ clusters. On the $Ga_{20}As_{19}H_{32}+O$ cluster one arsenic atom was replaced with an oxygen atom resulting in two highly polarized Ga-O bonds.

The above *brief* summary indicates that significant controversies about oxygen adsorption on gallium arsenide surface persist in the literature in spite of several studies. The central questions relate to possible adsorption sites and the nature of the GaAs surface upon adsorption. As a continuation of our studies of alkali and hydrogen adsorption on this surface [1], we present here a study of atomic oxygen adsorption on the GaAs surface, such surface being represented, as before, by a set of clusters. Specifically investigated are the adsorption sites, chemisorption energies, possibilities of charge transfers between the adatom and the Ga and the As atoms as also the highest occupied molecular orbital-lowest unoccupied molecular orbital (homo-lumo) gaps. We first comment on the computational methodology followed by results.

## II. Computational method and results

Both the unrestricted Hartree–Fock (UHF) theory and the many-body perturbation theory (MBPT) as used in this work are well documented in the literature [27-31]. Here we present only a basic equation to define some terms. In the MBPT, the energy is given by the linked diagram expansion:

$$\Delta E = E - E_0 = E_1 + E_{corr}$$
$$= \sum_{n=0}^{\infty} <\Phi_0 | V[(E_0 - H_0)^{-1}V]^n | \Phi_0>_L$$

(1)

where $\Phi_0$ is taken to be the unrestricted Hartree-Fock (UHF) wavefunction, $H_0$ is the sum of one-electron Fock operators, $E_0$ is the sum of the UHF orbital energies and $V = H - H_0$ perturbation, where $H$ is the usual electronic Hamiltonian. The subscript $L$ indicates the limitation to the linked diagrams. Though one can include various categories of infinite-order summations from Eq. (1), the method is usually limited by termination at some order of perturbation theory. In this work, because of severe demands on computational resources, we have carried out complete second-order calculations, which consist of all single and double-excitation terms for both the bare clusters and the chemisorbed systems.

One of the primary considerations involved in *ab initio* calculations is the type of basis set to be used [32]. Basis sets used in *ab initio* molecular-orbital computations usually involve some compromise between computational cost and accuracy. Keeping in mind the tremendous cost of *ab initio* calculations, specifically for large systems like gallium, and arsenic, we have elected to represent them by effective core potentials (ECP) or pseudopotentials (PP). In particular we have used the Hay-Wadt effective core potential (HWECP) and associated basis sets for gallium and arsenic [33], which are known to provide almost exact agreement with all electron results, to represent the GaAs clusters. To improve the accuracy of our calculations further, one *d* function was added to the HW basis sets. The exponent of the *d* function was chosen to provide the minimum energy for the $Ga_2$ and $As_2$ dimers, with the bond lengths fixed at experimental values [34]. The values of the exponents found were $d_{Ga}$ = 0.17 and $d_{As}$ = 0.28 [38]. For oxygen, we used Dunning's correlation consistent double-zeta contracted basis set [3s, 2p, 1d], augmented by three diffuse s, p, and d functions, respectively [35]. For the O atom, this basis set produced an electron affinity of 1.28eV and an ionization potential of 13.09eV. The corresponding experimental values are 1.46eV and 13.62eV, respectively. Obviously, this basis set for O can be considered as satisfactory for the purpose of our computations. All computations were done using a parallel version of GAUSSIAN 98 [36] on a Compaq Alpha ES40 parallel supercomputer at the University of Texas at Arlington.

In this work, we considered clusters representing two different reconstructed surfaces: the (2 × 1) and the *β* (4 × 2) surfaces [1]. Five different clusters were constructed, the smallest being the $Ga_4As_4H_{12}$ with two Ga atoms in the first layer and the largest being $Ga_{19}As_{15}H_{39}$, with each cluster constructed with Ga and As atoms located at the bulk lattice sites given by the zinc-blende structure with an experimental lattice constant of 5.654Å. Ga atoms terminated the first or the top layer and the second layer was composed of As atoms while the third layer was composed of Ga atoms. The cluster sizes increased in both transverse dimensions as well as number of layers, the maximum being three. Hydrogen atoms were used to saturate the dangling bonds, except above the surface, at an energy optimized bond length of 1.511Å, which is in agreement with the work of Nonoyama *et al*. [37] who used a similar approach for constructing $Ga_4As_4H_{12}$ for chemisorption of atomic and molecular hydrogen on GaAs (100). Due to severe demands on computational resources, total energy optimization was carried out only for the smallest cluster, $Ga_4As_4H_{12}$, by allowing dimerization of the surface Ga atoms. From this process, the surface Ga-Ga dimer bond length was found to be 2.758Å. This length was then used for the $Ga_5As_6H_{16}$, $Ga_7As_6H_{16}$, $Ga_7As_6H_{19}$, and $Ga_{19}As_{15}H_{39}$ clusters. Specifically, the $Ga_4As_4H_{12}$, $Ga_5As_6H_{16}$, $Ga_7As_6H_{16}$, and the $Ga_7As_6H_{19}$ clusters represent the (2x1) surface and the $Ga_{19}As_{15}H_{39}$ cluster represents the *β* (4 × 2) surface [1]. Different sizes of clusters are used to represent the same surface because of non-uniqueness of a specific cluster to represent a surface and also to study dependence and convergence of cluster properties with respect to cluster sizes. As a comparison, Guo-Ping and Ruda [34] used a surface Ga-Ga dimer bond length of 2.80Å in a similar *ab initio* cluster study of the adsorption of sulfur on the Ga-rich GaAs(100) surface. The total energies and binding energies of all the clusters at the UHF and the MP2 levels are shown in Table 1. The binding energies per atom were calculated from

$$E_b = (xE(Ga) + yE(As) + zE(H) – E(Ga_xAs_yH_z)) / (x + y + z). \quad (2)$$

We note that the binding energies oscillate with the number of atoms in the clusters at both levels of theory and the binding energies at the MP2 level of theory are higher than the corresponding energies at the UHF level of theory. It is known that correlation effects typically increase the binding or cohesive energy in a cluster.

To study atomic oxygen adsorption on the Ga-rich GaAs(100) reconstructed surface, we considered six adsorption sites of high symmetry, four surface sites and two interstitial sites, namely cage and trough. The sites yielding the highest chemisorption energies are shown in figures 1-6. . All sites were chosen because of their associated point symmetries. The top, bridge, and the interstitial sites were chosen for their $\sigma_V$ inversion symmetry through a plane perpendicular to the surface dimer mid-point. The cave, hollow, and trough sites were chosen for their $C_{4V}$ point rotational symmetry about an axis normal to the (100) plane. In the top site, the adatom is allowed to approach a path directly on top of a Ga atom, whereas in the interstitial sites, the adatom migrated inside the cluster. To examine the relative stability of chemisorption at the different sites, the chemisorption energies are calculated from,

$$E_C = E(O) + E(Ga_xAs_yH_z) - E(O+Ga_xAs_yH_z). \quad (3)$$

For all surface sites, the height of the adatom above the top Ga layer was varied to yield the maximum chemisorption energy (*i.e.* a minimum of the $E_c$ versus d curve, with the sign of the $E_c$ changed). Typically, about ten data points were generated to get accurate values of the O adatom distance and the chemisorption energy. For the sake of brevity, we have elected not to show all the figures, though figures 7-14 show some of the $E_c$ versus d curves.

The numerical results at the unrestricted Hartree-Fock and second-order many-body perturbation theory levels, along with our previous results for H chemisorption on the GaAs surface [1], are shown in table 2. We note that for all sites, MP2 theory predicts higher chemisorption energies compared to the UHF results. Also, though UHF theory predicts several sites to have negative chemisorption energies, all sites appear to be potential sites for atomic oxygen chemisorption at the higher MP2 level of theory due to the positive values of the chemisorption energies. In fact, the chemisorption energies are high, varying from 1.633eV to 6.347eV. The dimerized bridge sites are most favored, with the $O+Ga_{19}As_{15}H_{39}$ site having the highest chemisorption energy, namely 6.347eV, of all the sites considered. The second most preferred site is also a bridge site, corresponding to the $Ga_7As_6H_{19}$ cluster. This is in agreement with the work of Stroscio *et al.* [6], who found that atomic oxygen is bonded in an interchain bridging position on the GaAs (110) surface. Taguchi and Kageshima [19] also observed that the atomic configuration in GaAs is Ga-O-Ga, consistent with a bridging position. If one compares O chemisorption with H chemisorption on the GaAs(100) surface, we note first that, for H chemisorption, the top site was the most preferred site, with a chemisorption energy of 2.844eV. Also, O interaction with the GaAs surface appears to be much stronger, with chemisorption energies significantly higher in all cases. Even at the correlated level of theory, H does not chemisorb at several sites, as compared to oxygen which, as indicated before, chemisorbs at all sites. It is worth noting here that the H chemisorption studies were performed at the restricted open-shell Hartree-Fock theory (ROMP2) level. However, we do not believe that this will, in any way, alter the main conclusions noted above. Table 3 lists the adatom to the nearest surface neighbor bond length in Å for the various cluster sites. The distances vary significantly for the different sites from 1.720Å to 3.141Å, with the adatom being at a distance of 1.829Å from the nearest neighbor surface atom for the most preferred bridge site. The largest distances of the O adatom are found for the hollow and cave sites for the (2 × 1) and β(4 × 2) clusters and no correlation is observed between the adatom bond lengths and the chemisorption energies. As a comparison, for the H adatom chemisorption, the distances vary from 1.585Å to 3.593Å, the largest distance being for the hollow site corresponding to the O+$Ga_{19}As_{15}H_{39}$ cluster, followed by the hollow site for the $Ga_7As_6H_{19}$ cluster.

We also studied the effects on the homo-lumo (highest occupied molecular orbital-lowest unoccupied molecular orbital) gap of the GaAs(100) surface due to oxygen adsorption. Table 4 lists the results of our study. We note that in eleven of the seventeen clusters studied, the gaps increase in value, from 0.201eV for O adsorption on the $Ga_7As_6H_{15}$ cluster in the cave site to 4.206eV for O adsorption on the $Ga_7As_6H_{19}$ cluster in the top site suggesting a transition to a semi-insulating state for the GaAs surface, consistent with experimental data. For four cases of adsorption on the $Ga_4As_4H_{12}$ cluster and two cases for the $Ga_{19}As_{15}H_{39}$ cluster, the gaps actually decrease in value. Five out of these six cases, the chemisorption energies are rather high, from 4.481eV to 6.347eV, indicating an inverse correlation between chemisorption and the change in homo-lumo gap. The general trend noted here is in some contrast to our published results on alkali atom chemisorption on Ga-rich GaAs(100) and (110) surfaces [1,38], where the gap, in general, decreased in value, indicating possibilities of metallization of the substrate.

We have also carried out an analysis of the atomic charge distributions using Mulliken population analysis [39]. For the sake of brevity, we have shown the results in Tables 5-8 only for the most stable sites in each adsorption category, namely the top and cage sites for the $Ga_4As_4H_{12}$ cluster, trough site for $Ga_5As_6H_{16}$ cluster, cave site for the $Ga_7As_6H_{16}$ cluster, and bridge and hollow sites for the $Ga_{19}As_{15}H_{39}$ cluster. We also show the charge distribution plots only for some selected cases. In all cases, the oxygen adatom gains negative charge as expected due to the relative electronegativities of the surrounding atoms in the clusters. The charge gained by oxygen ranges from 1.108$e$ on the 1b top site of the $Ga_4As_4H_{12}$ + O cluster, to 0.109$e$ at the interstitial trough site of the $Ga_5As_6H_{16}$ + O cluster. For the 1b top and cage sites of the $Ga_4As_4H_{12}$ + O cluster, the first layer Ga atoms lose a significant amount of charge whereas the charge transfer, loss or gain, for the second layer As and third layer Ga atoms are lower (Table 5). This suggests that the surface Ga-O interaction plays a significant role in chemisorption for this cluster. Table 6 lists the atomic charge distributions for the $Ga_5As_6H_{16}$ bare and $Ga_5As_6H_{16}$ + O chemisorbed trough site. With 0.109$e$ of charge transfer to the oxygen adatom this site yields the least amount of charge transfer to the adatom. For the chemisorbed cluster, surface Ga atoms gain negative charge and there is minimal charge transfer for the second layer As atoms. The two third layer Ga atoms involved in bonding with the oxygen gain a significant amount of negative charge suggesting that the interstitial Ga atoms play a significant role in chemisorption for the cluster. This is in agreement with the work of Landgren *et al.* [3] but in contrast to the work of Lee *et al.* [16] which suggested that subsurface oxidation does not take place. Figures 15 and 16 [40] show the electronic charge distribution through one plane for the $Ga_5As_6H_{16}$ bare and $Ga_5As_6H_{16}$ + O chemisorbed trough site. Atomic charge distributions for the $Ga_7As_6H_{16}$ (2 × 1) surface are shown in table 7 and figure 17 shows the charge distribution for the second layer of the bare cluster. For the $Ga_7As_6H_{16}$ + O cave site, the oxygen adatom gains 0.702$e$ of charge. The surface Ga atoms all lose charge while the results for the second layer As are mixed. All three third layer Ga atoms in the chemisorbed cluster gain a small amount of negative charge. Table 8 lists the atomic charge distributions for the $Ga_{19}As_{15}H_{39}$ cluster representing the ß(4 × 2) surface. The oxygen adatom gains 0.916$e$ of charge for the hollow site and 0.899$e$ of charge for the bridge site. There appears to be an inverse correlation between the negative charge transfer to the oxygen adatom and the chemisorption energy for this cluster as the bridge site clearly has the higher $E_C$. The results for the first layer Ga, second layer As and third layer Ga are mixed with some atoms gaining negative charge and some losing negative charge. Figures 18 and 19 show the electron charge densities for the $Ga_{19}As_{15}H_{39}$ bare and $Ga_{19}As_{15}H_{39}$ + O chemisorbed bridge sites. Figures 20 and 21 show the electron charge distribution through a plane for the $Ga_{19}As_{15}H_{39}$ bare and $Ga_{19}As_{15}H_{39}$ + O hollow site. For all clusters considered, the electronic charge densities are all redistributed such that the oxygen adatom attracts the bulk of the charge. This is true for both surface and interstitial sites suggesting that oxidation occurs not only on the surface but beneath the surface as well.

In summary**, w**e have carried out *ab initio* cluster calculations to study oxygen chemisorption on the Ga-rich GaAs(100) surface. Out of the six high-symmetry sites, all six favored chemisorption with the $Ga_{19}As_{15}H_{39}$ dimerized bridge site having the highest chemisorption energy followed by the $Ga_4As_4H_{12}$ and $Ga_7As_6H_{19}$ dimerized bridge sites. In both trough and cage sites, the O adatom was found to favor an interstitial position. This is in agreement with recent studies that suggest that the Ga-O-Ga bond of the oxygen defect is most favorable. The distance of the adatom from the nearest surface neighbor was found to lie between 1.7 to 3.1Å and the homo-lumo gap was found to vary between 1.6 to 6.4eV.

Finally, the authors gratefully acknowledge partial support from the Welch Foundation, Houston, Texas (Grant No. Y-1525). We also would like to thank Dr. Ronald Schailey for invaluable discussions and encouragement during the early phases of this work.

Table 1: Total energy (a.u.) and binding energy (eV) of the (2 × 1) and β(4 × 2) bare clusters.

| Cluster | $E_{tot}$(UHF) | $E_{tot}$(UMP2) | $E_b$(UHF) | $E_b$(UMP2) |
|---|---|---|---|---|
| $Ga_4As_4H_{12}$ | -38.653 | -39.355 | 1.444 | 1.944 |
| $Ga_5As_6H_{16}$ | -54.724 | -55.749 | 1.290 | 1.847 |
| $Ga_7As_6H_{16}$ | -58.817 | -59.947 | 1.386 | 1.950 |
| $Ga_7As_6H_{19}$ | -60.576 | -61.713 | 1.481 | 1.998 |
| $Ga_{19}As_{15}H_{39}$ | -149.264 | -152.236 | 1.301 | 1.900 |

Table 2: Chemisorption energy (eV) vs. cluster size and symmetry. $E_C$ (ROMP2) values are for hydrogen adsorption.

| Sites | Symmetry | Cluster | $E_C$(UHF) | $E_C$(UMP2) | $E_C$(ROMP2) |
|---|---|---|---|---|---|
| 1 (Top) | | | | | |
| 1a | 2 × 1 | O + $Ga_4As_4H_{12}$ | 0.522 | 4.481 | 2.786 |
| 1b | 2 × 1 | O + $Ga_4As_4H_{12}$ | 0.549 | 4.482 | |
| 1a | 2 × 1 | O + $Ga_7As_6H_{19}$ | 2.122 | 3.602 | 2.844 |
| 1b | 2 × 1 | O + $Ga_7As_6H_{19}$ | 2.196 | 3.626 | |
| 1b | 4 × 2 | O + $Ga_{19}As_{15}H_{39}$ | 2.543 | 4.355 | |
| 2 (Bridge) | 2 × 1 | O + $Ga_4As_4H_{12}$ | 0.558 | 4.602 | 0.422 |
| | 2 × 1 | O + $Ga_7As_6H_{19}$ | 2.838 | 5.758 | 1.214 |
| | 4 × 2 | O + $Ga_{19}As_{15}H_{39}$ | 4.204 | 6.347 | |
| 3 (Hollow) | 2 × 1 | O + $Ga_7As_6H_{19}$ | -2.316 | 2.824 | 0.797 |
| | 4 × 2 | O + $Ga_{19}As_{15}H_{39}$ | -2.071 | 3.408 | |
| 4 (Cave) | 2 × 1 | O + $Ga_7As_6H_{16}$ | -1.216 | 3.692 | -0.144 |
| | 4 × 2 | O + $Ga_{19}As_{15}H_{39}$ | -2.499 | 2.758 | |
| 5 (Trough) | 2 × 1 | O + $Ga_5As_6H_{16}$ | -0.129 | 2.377 | -0.938 |
| 5a | 4 × 2 | O + $Ga_{19}As_{15}H_{39}$ | -1.983 | 1.633 | |
| 5b | 4 × 2 | O + $Ga_{19}As_{15}H_{39}$ | -1.744 | 2.229 | |
| 6 (Cage) | 2 × 1 | O + $Ga_4As_4H_{12}$ | 1.525 | 5.635 | 1.999 |
| | 2 × 1 | O + $Ga_7As_6H_{19}$ | 1.262 | 4.303 | 2.113 |

Table 3: Bond Length (Å) of adatom vs. cluster size and symmetry.

| Sites | Symmetry | Cluster | Adatom-nearest surface neighbor bond length | Adatom-nearest surface neighbor bond length for H |
|---|---|---|---|---|
| 1 (Top) | | | | |
| 1a | 2 × 1 | O + $Ga_4As_4H_{12}$ | 1.720 | 1.579 |
| 1b | 2 × 1 | O + $Ga_4As_4H_{12}$ | 1.720 | |
| 1a | 2 × 1 | O + $Ga_7As_6H_{19}$ | 1.900 | 1.585 |
| 1b | 2 × 1 | O + $Ga_7As_6H_{19}$ | 1.900 | |
| 1b | 4 × 2 | O + $Ga_{19}As_{15}H_{39}$ | 1.700 | 1.763 |
| 2 (Bridge) | 2 × 1 | O + $Ga_4As_4H_{12}$ | 1.895 | 1.731 |
| | 2 × 1 | O + $Ga_7As_6H_{19}$ | 1.829 | 1.742 |
| | 4 × 2 | O + $Ga_{19}As_{15}H_{39}$ | 1.829 | 1.850 |
| 3 (Hollow) | 2 × 1 | O + $Ga_7As_6H_{19}$ | 3.141 | 3.501 |
| | 4 × 2 | O + $Ga_{19}As_{15}H_{39}$ | 2.619 | 3.593 |
| 4 (Cave) | 2 × 1 | O + $Ga_7As_6H_{16}$ | 2.512 | 2.622 |
| | 4 × 2 | O + $Ga_{19}As_{15}H_{39}$ | 2.753 | 2.938 |
| 5 (Trough) | 2 × 1 | O + $Ga_5As_6H_{16}$ | 2.009 | 2.985 |
| 5a | 4 × 2 | O + $Ga_{19}As_{15}H_{39}$ | 1.999 | |
| 5b | 4 × 2 | O + $Ga_{19}As_{15}H_{39}$ | 1.999 | 2.925 |
| 6 (Cage) | 2 × 1 | O + $Ga_4As_4H_{12}$ | 1.824 | 1.818 |
| | 2 × 1 | O + $Ga_7As_6H_{19}$ | 1.746 | 1.765 |

Table 4: HOMO-LUMO gap (eV) vs. cluster size and symmetry.

| Sites | Symmetry | Cluster | Gap | Cluster | Gap | ΔGap |
|---|---|---|---|---|---|---|
| 1 (Top) | | | | | | |
| 1a | 2 × 1 | $Ga_4As_4H_{12}$ | 7.462 | $O + Ga_4As_4H_{12}$ | 6.439 | -1.023 |
| 1b | 2 × 1 | $Ga_4As_4H_{12}$ | 7.462 | $O + Ga_4As_4H_{12}$ | 6.435 | -1.027 |
| 1a | 2 × 1 | $Ga_7As_6H_{19}$ | 2.057 | $O + Ga_7As_6H_{19}$ | 6.263 | 4.206 |
| 1b | 2 × 1 | $Ga_7As_6H_{19}$ | 2.057 | $O + Ga_7As_6H_{19}$ | 2.995 | 0.938 |
| 1b | 4 × 2 | $Ga_{19}As_{15}H_{39}$ | 2.385 | $O + Ga_{19}As_{15}H_{39}$ | 4.096 | 1.711 |
| 2 (Bridge) | 2 × 1 | $Ga_4As_4H_{12}$ | 7.462 | $O + Ga_4As_4H_{12}$ | 5.263 | -2.199 |
|  | 2 × 1 | $Ga_7As_6H_{19}$ | 2.057 | $O + Ga_7As_6H_{19}$ | 6.005 | 3.948 |
|  | 4 × 2 | $Ga_{19}As_{15}H_{39}$ | 2.385 | $O + Ga_{19}As_{15}H_{39}$ | 2.247 | -0.138 |
| 3 (Hollow) | 2 × 1 | $Ga_7As_6H_{19}$ | 2.057 | $O + Ga_7As_6H_{19}$ | 5.447 | 3.390 |
|  | 4 × 2 | $Ga_{19}As_{15}H_{39}$ | 2.385 | $O + Ga_{19}As_{15}H_{39}$ | 2.695 | 0.310 |
| 4 (Cave) | 2 × 1 | $Ga_7As_6H_{16}$ | 3.698 | $O + Ga_7As_6H_{16}$ | 3.899 | 0.201 |
|  | 4 × 2 | $Ga_{19}As_{15}H_{39}$ | 2.385 | $O + Ga_{19}As_{15}H_{39}$ | 1.643 | -0.742 |
| 5 (Trough) | 2 × 1 | $Ga_5As_6H_{16}$ | 4.584 | $O + Ga_5As_6H_{16}$ | 5.235 | 0.651 |
| 5a | 4 × 2 | $Ga_{19}As_{15}H_{39}$ | 2.385 | $O + Ga_{19}As_{15}H_{39}$ | 2.881 | 0.496 |
| 5b | 4 × 2 | $Ga_{19}As_{15}H_{39}$ | 2.385 | $O + Ga_{19}As_{15}H_{39}$ | 3.405 | 1.020 |
| 6 (Cage) | 2 × 1 | $Ga_4As_4H_{12}$ | 7.462 | $O + Ga_4As_4H_{12}$ | 6.100 | -1.362 |
|  | 2 × 1 | $Ga_7As_6H_{19}$ | 2.057 | $O + Ga_7As_6H_{19}$ | 5.321 | 3.264 |

Table 5: Atomic charge distributions for the $Ga_4As_4H_{12}$ (2 × 1) surface.

| Layer | Atom | Free Cluster | Top Site 1b | Cage Site |
|---|---|---|---|---|
| Adatom | O |  | -1.108 | -0.227 |
| 1st | Ga | -0.006 | 0.220 | 0.248 |
| 1st | Ga | -0.006 | 0.927 | 0.249 |
| 2nd | As | -0.088 | -0.095 | -0.090 |
| 2nd | As | -0.088 | -0.116 | -0.091 |
| 2nd | As | -0.088 | -0.116 | -0.091 |
| 2nd | As | -0.088 | -0.095 | -0.090 |
| 3rd | Ga | 0.204 | 0.162 | 0.140 |
| 3rd | Ga | 0.204 | 0.162 | 0.140 |

Table 6: Atomic charge distributions for the $Ga_5As_6H_{16}$ (2 × 1) surface.

| Layer | Atom | Free Cluster | Trough Site |
|---|---|---|---|
| Adatom | O |  | -0.109 |
| 1st | Ga | 0.222 | 0.129 |
| 1st | Ga | 0.222 | 0.129 |
| 2nd | As | -0.245 | -0.196 |
| 2nd | As | -0.196 | -0.263 |
| 2nd | As | -0.246 | -0.196 |
| 2nd | As | -0.196 | -0.264 |
| 2nd | As | -0.060 | -0.089 |
| 2nd | As | -0.060 | -0.089 |
| 3rd | Ga | 0.206 | 0.036 |
| 3rd | Ga | 0.196 | 0.203 |
| 3rd | Ga | 0.319 | 0.018 |

Table 7: Atomic charge distributions for the $Ga_7As_6H_{16}$ (2 × 1) surface.

| Layer | Atom | Free Cluster | Cave Site |
|---|---|---|---|
| Adatom | O |  | -0.702 |
| 1st | Ga | 0.105 | 0.312 |
| 1st | Ga | 0.105 | 0.313 |
| 1st | Ga | 0.106 | 0.307 |
| 1st | Ga | 0.106 | 0.307 |
| 2nd | As | -0.093 | -0.094 |
| 2nd | As | -0.244 | -0.156 |
| 2nd | As | -0.093 | -0.095 |
| 2nd | As | -0.244 | -0.157 |
| 2nd | As | -0.093 | -0.096 |
| 2nd | As | -0.093 | -0.096 |
| 3rd | Ga | 0.168 | 0.136 |
| 3rd | Ga | 0.168 | 0.136 |
| 3rd | Ga | 0.283 | 0.274 |

Table 8: Atomic charge distributions for the $Ga_{19}As_{15}H_{39}$ $\beta(4 \times 2)$ surface.

| Layer | Atom | Free Cluster | Bridge Site | Hollow Site |
|---|---|---|---|---|
| Adatom | O |  | -0.899 | -0.916 |
| 1st | Ga | 0.122 | 0.153 | 0.169 |
| 1st | Ga | 0.056 | 0.207 | 0.213 |
| 1st | Ga | 0.294 | 0.167 | 0.220 |
| 1st | Ga | 0.244 | 0.788 | 0.218 |
| 1st | Ga | 0.188 | 0.735 | 0.398 |
| 1st | Ga | 0.319 | 0.290 | 0.404 |
| 1st | Ga | 0.270 | 0.280 | 0.258 |
| 1st | Ga | 0.284 | 0.303 | 0.237 |
| 1st | Ga | -0.011 | -0.015 | 0.294 |
| 2nd | As | -0.171 | -0.186 | -0.205 |
| 2nd | As | -0.179 | -0.169 | -0.147 |
| 2nd | As | -0.278 | -0.327 | -0.198 |
| 2nd | As | -0.356 | -0.113 | -0.259 |
| 2nd | As | -0.299 | -0.315 | -0.297 |
| 2nd | As | -0.077 | -0.127 | -0.090 |
| 2nd | As | -0.289 | -0.307 | -0.290 |
| 2nd | As | 0.021 | -0.191 | -0.135 |
| 2nd | As | -0.145 | -0.141 | -0.098 |
| 2nd | As | -0.212 | -0.238 | -0.225 |
| 2nd | As | -0.303 | -0.323 | -0.193 |
| 2nd | As | -0.196 | -0.213 | -0.109 |
| 2nd | As | -0.075 | -0.075 | -0.079 |
| 2nd | As | -0.194 | -0.190 | -0.213 |
| 2nd | As | -0.079 | -0.081 | -0.047 |
| 3rd | Ga | 0.291 | 0.260 | 0.303 |
| 3rd | Ga | 0.190 | 0.200 | 0.217 |
| 3rd | Ga | 0.300 | 0.263 | 0.217 |
| 3rd | Ga | 0.255 | 0.252 | 0.348 |
| 3rd | Ga | 0.226 | 0.231 | 0.181 |
| 3rd | Ga | 0.214 | 0.195 | 0.202 |
| 3rd | Ga | 0.275 | 0.247 | 0.255 |
| 3rd | Ga | 0.256 | 0.245 | 0.298 |
| 3rd | Ga | 0.184 | 0.184 | 0.182 |
| 3rd | Ga | 0.177 | 0.175 | 0.173 |

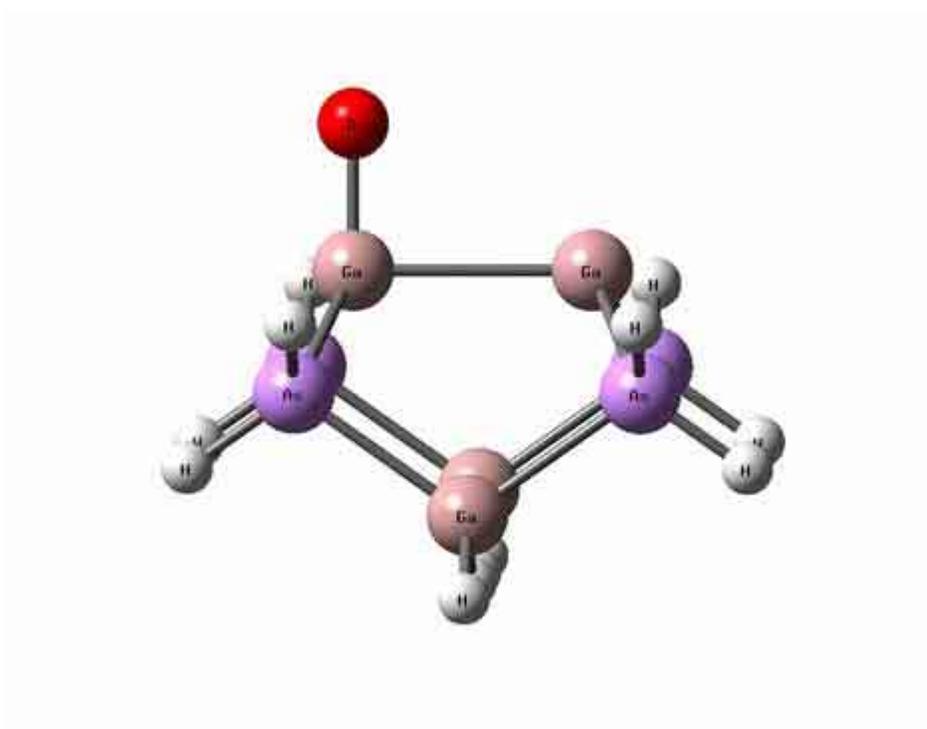

Fig. 1. $Ga_4As_4H_{12}$+O Site 1b.

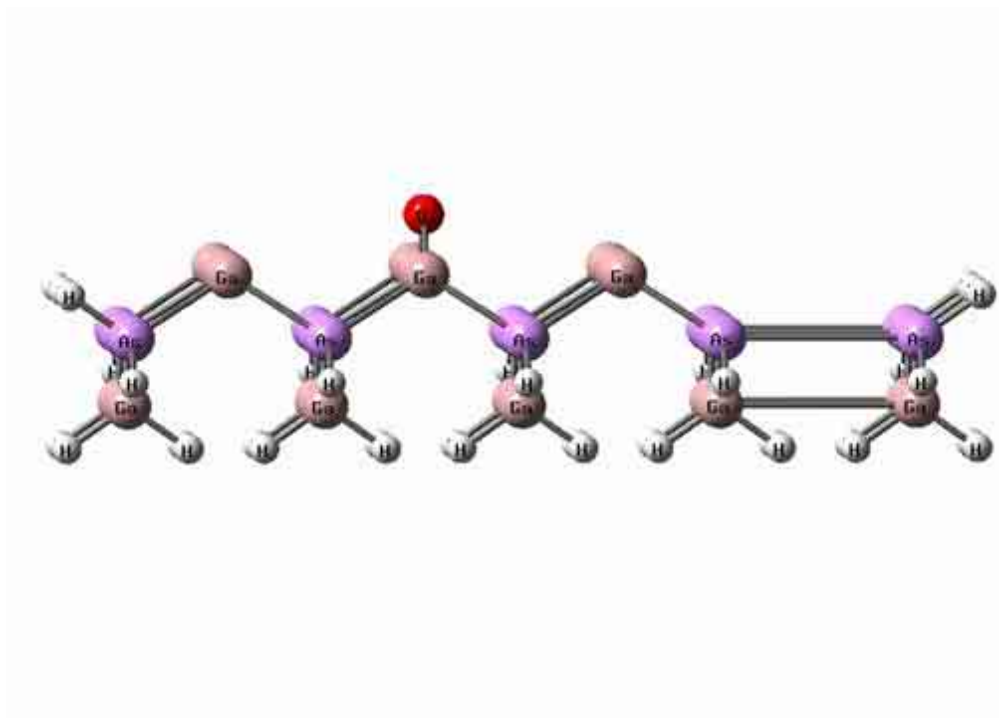

Fig. 2. $Ga_{19}As_{15}H_{39}$ + O Bridge Site.

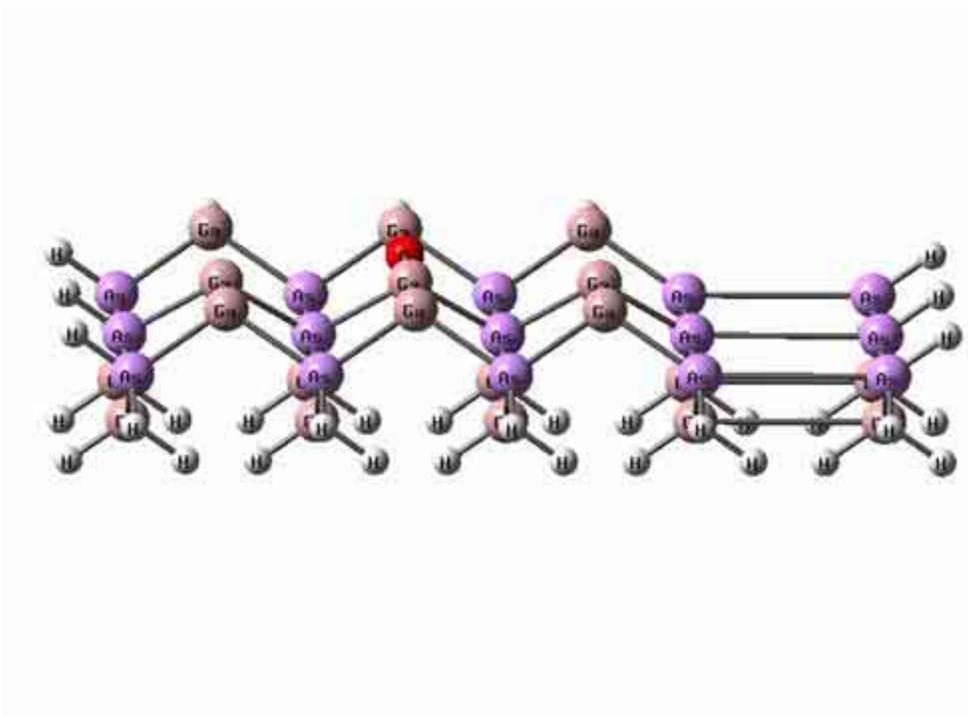

Fig. 3. $Ga_{19}As_{15}H_{39}$ + O Hollow Site.

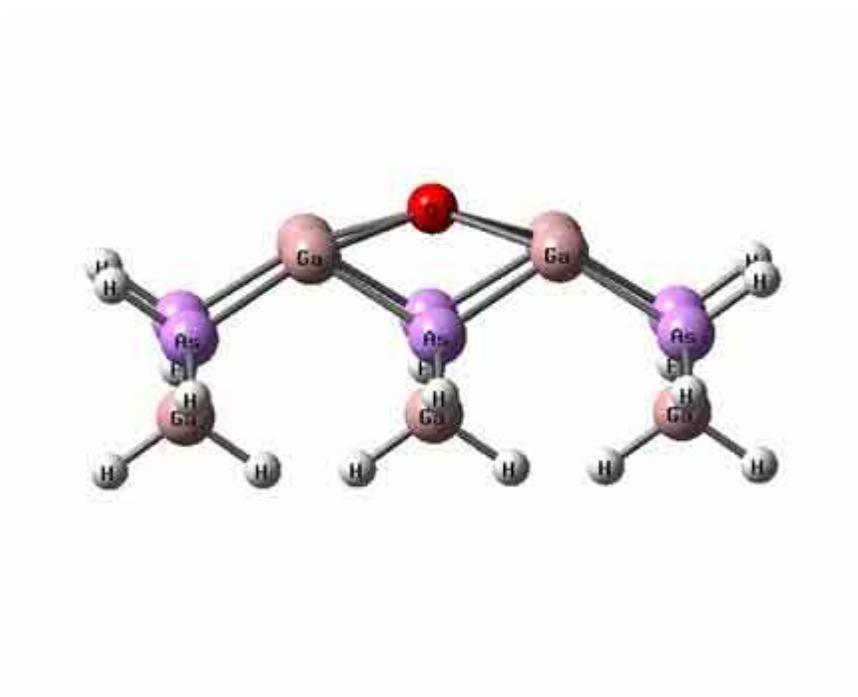

Fig. 4. $Ga_7As_6H_{16}$ + O Cave Site.

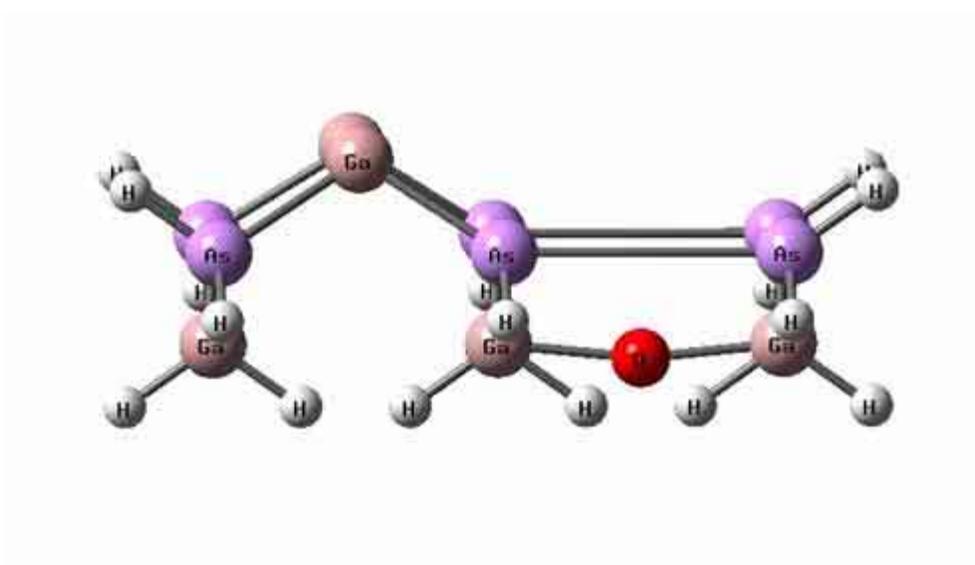

Fig. 5. $Ga_5As_6H_{16}$ + O Trough Site.

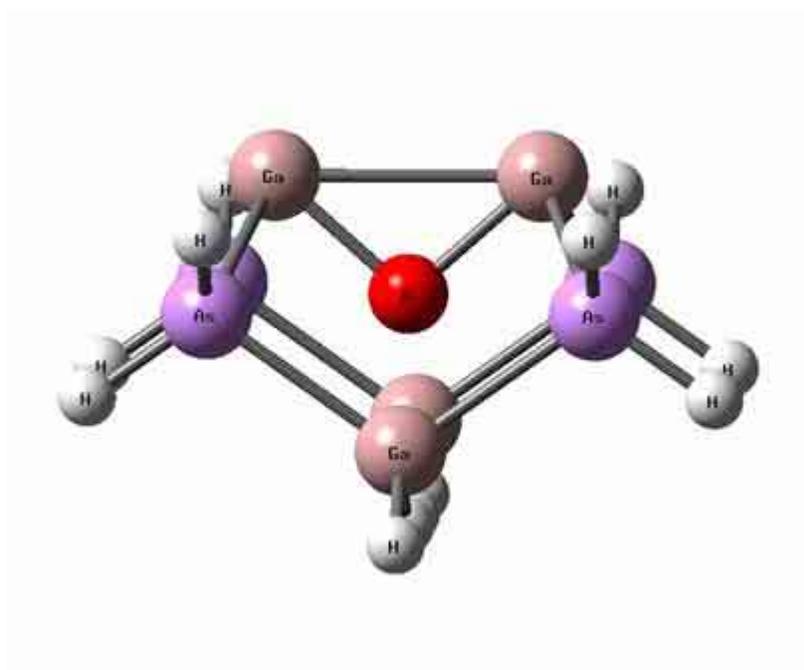

Fig. 6. $Ga_4As_4H_{12}$ + O Cage Site.

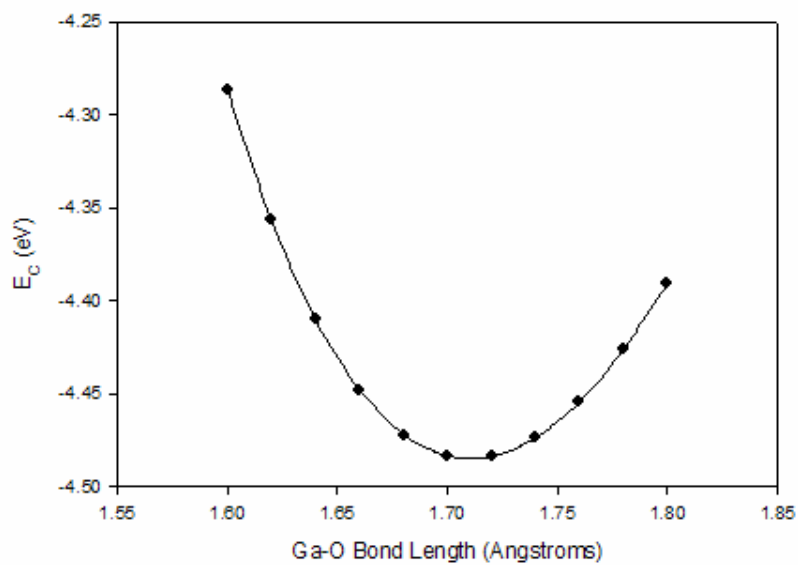

Fig. 7. Chemisorption energy $E_C$ (eV) vs. Ga-O bond length (Å) for $Ga_4As_4H_{12}$ + O top site.

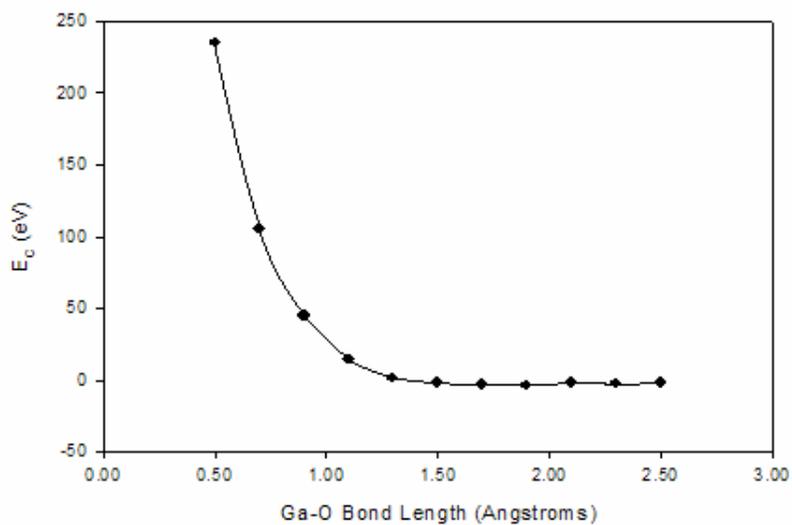

Fig. 8. Chemisorption energy $E_C$ (eV) vs. Ga-O bond length (Å) for $Ga_7As_6H_{19}$ + O top site.

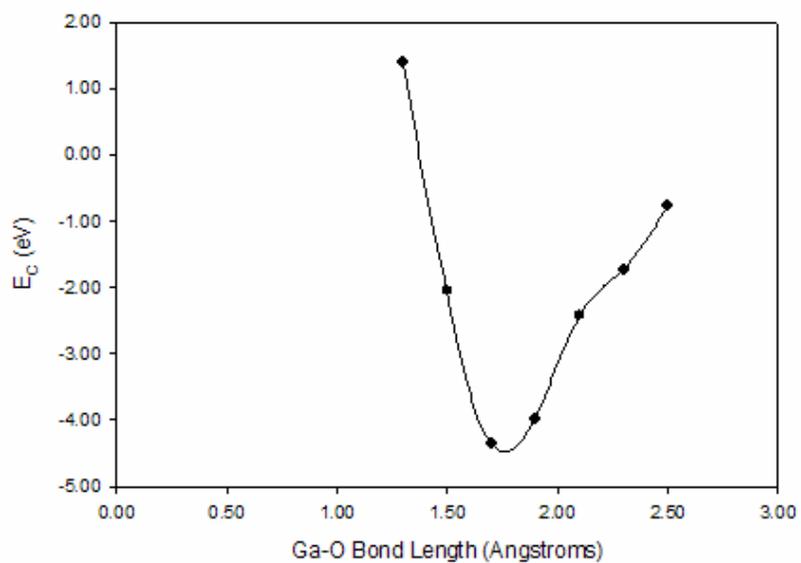

Fig. 9. Chemisorption energy $E_C$ (eV) vs. Ga-O bond length (Å) for $Ga_{19}As_{15}H_{39}$ + O top site.

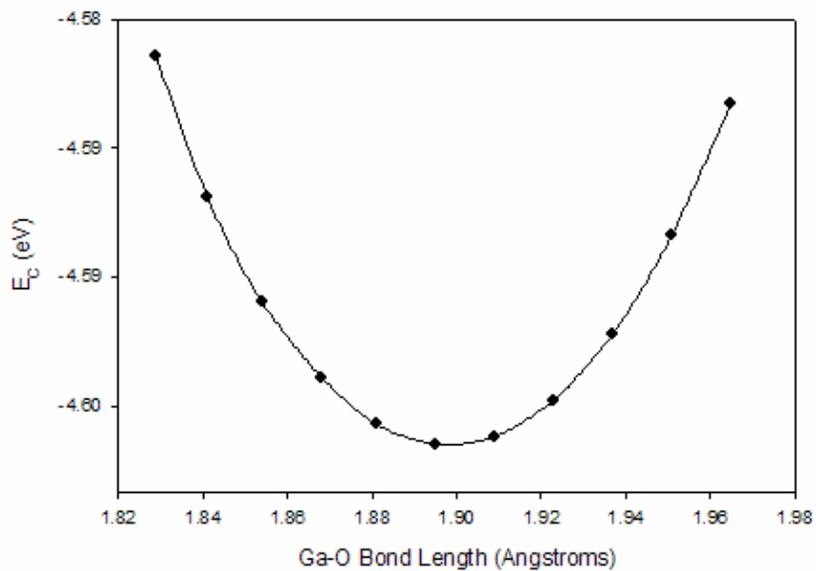

Fig. 10. Chemisorption energy $E_C$ (eV) vs. Ga-O bond length (Å) for $Ga_4As_4H_{12}$ + O bridge site.

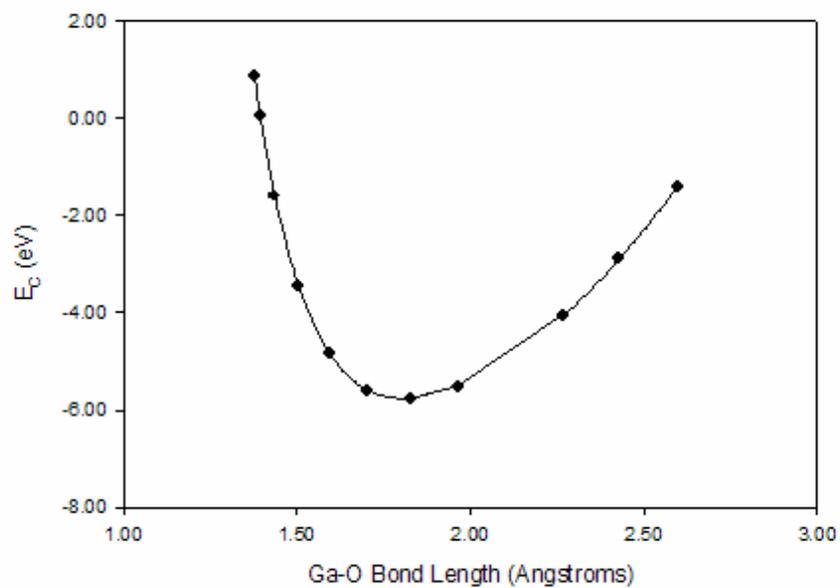

Fig. 11. Chemisorption energy $E_C$ (eV) vs. Ga-O bond length (Å) for $Ga_7As_6H_{19}$ + O bridge site.

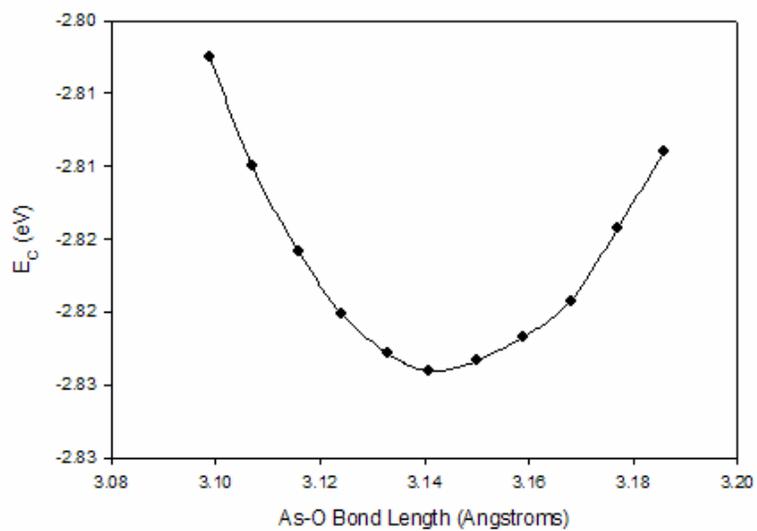

Fig. 12. Chemisorption energy $E_C$ vs. As-O bond length (Å) for $Ga_7As_6H_{19}$ + O hollow site.

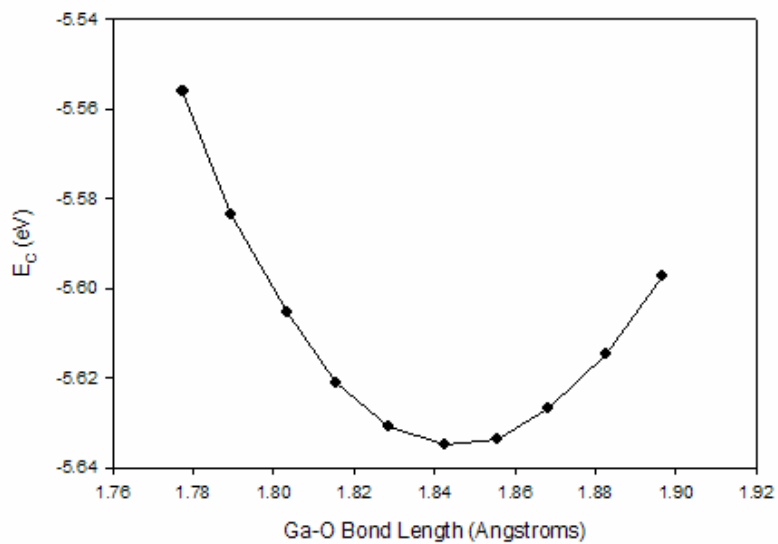

Fig. 13. Chemisorption energy $E_C$ vs. (eV) Ga-O bond length (Å) for $Ga_4As_4H_{12}$ + O cage site.

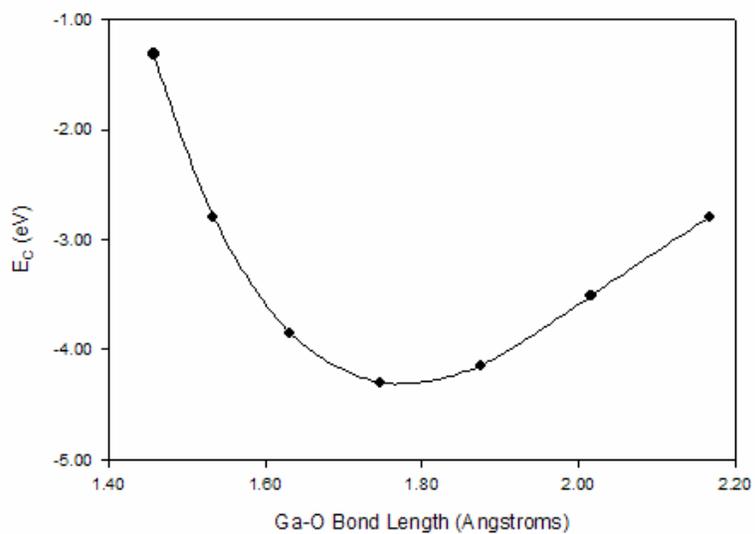

Fig. 14. Chemisorption energy $E_C$ (eV) vs. Ga-O bond length (Å) for $Ga_7As_6H_{19}$ + O cage site.

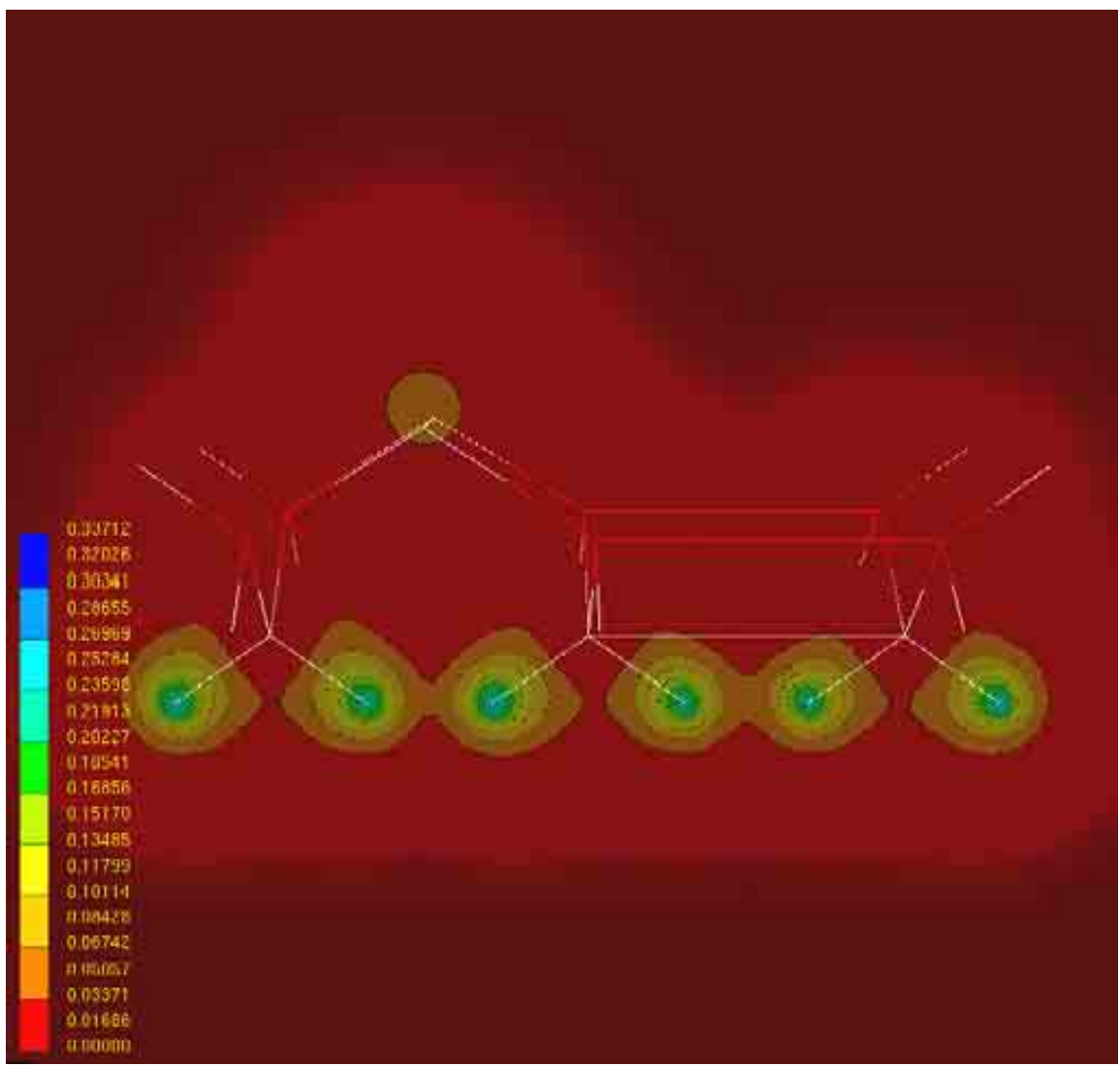

Figure 15. Electron charge distribution through a plane for the $Ga_5As_6H_{16}$ bare cluster.

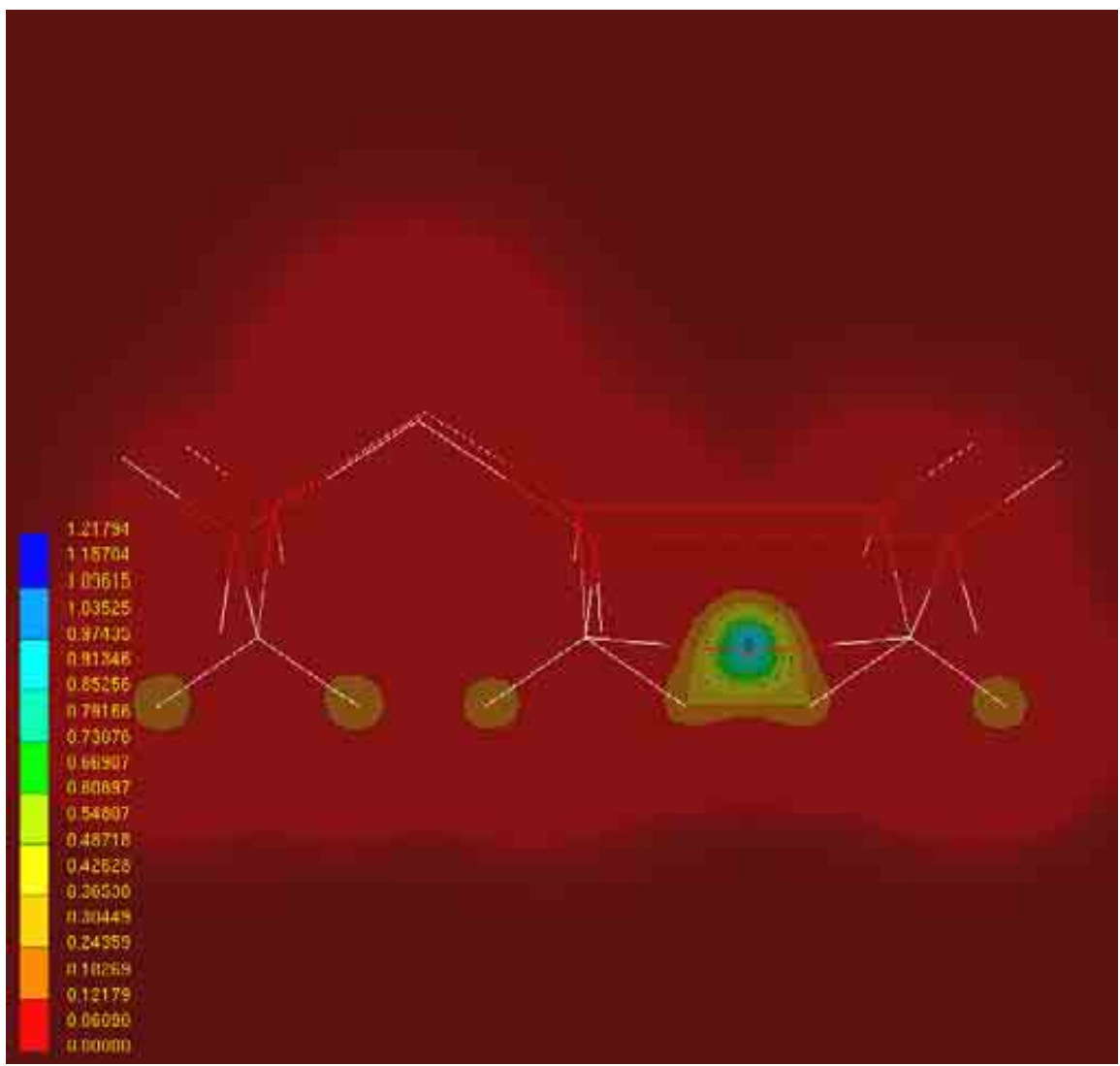

Figure 16. Electron charge distribution through a plane for the $Ga_5As_6H_{16}$ + O trough site.

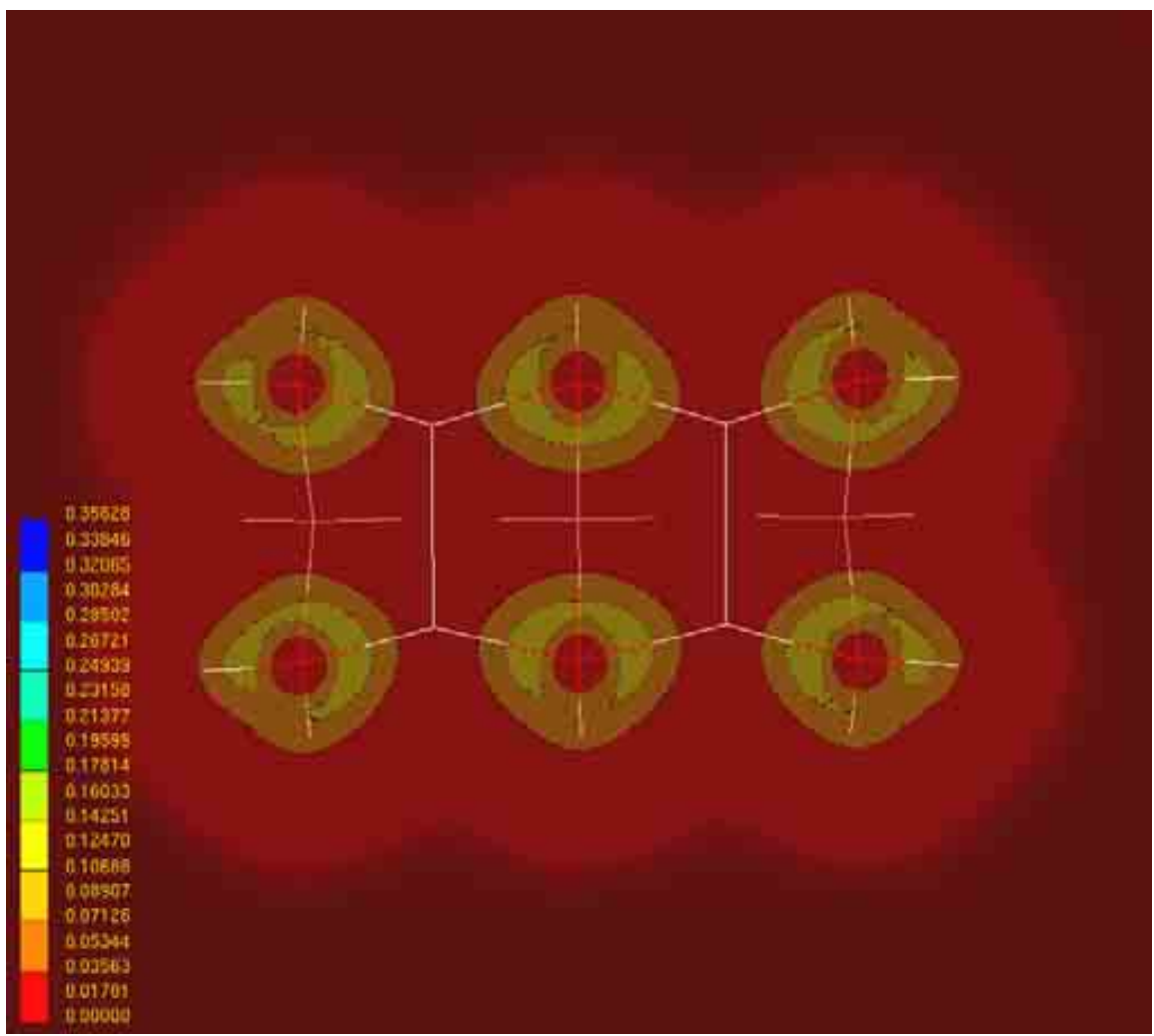

Figure 17. Electron charge distribution for the second layer of the $Ga_7As_6H_{16}$ bare cluster.

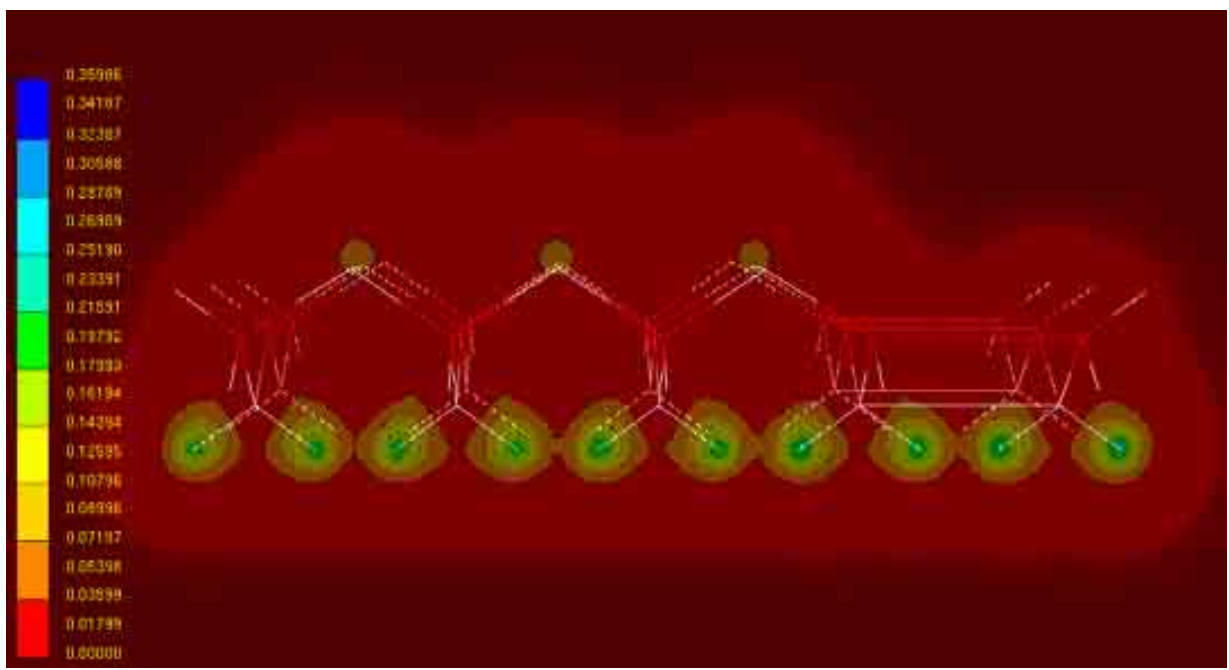

Figure 18. Electron charge distribution through a plane for the $Ga_{19}As_{15}H_{39}$ bare cluster.

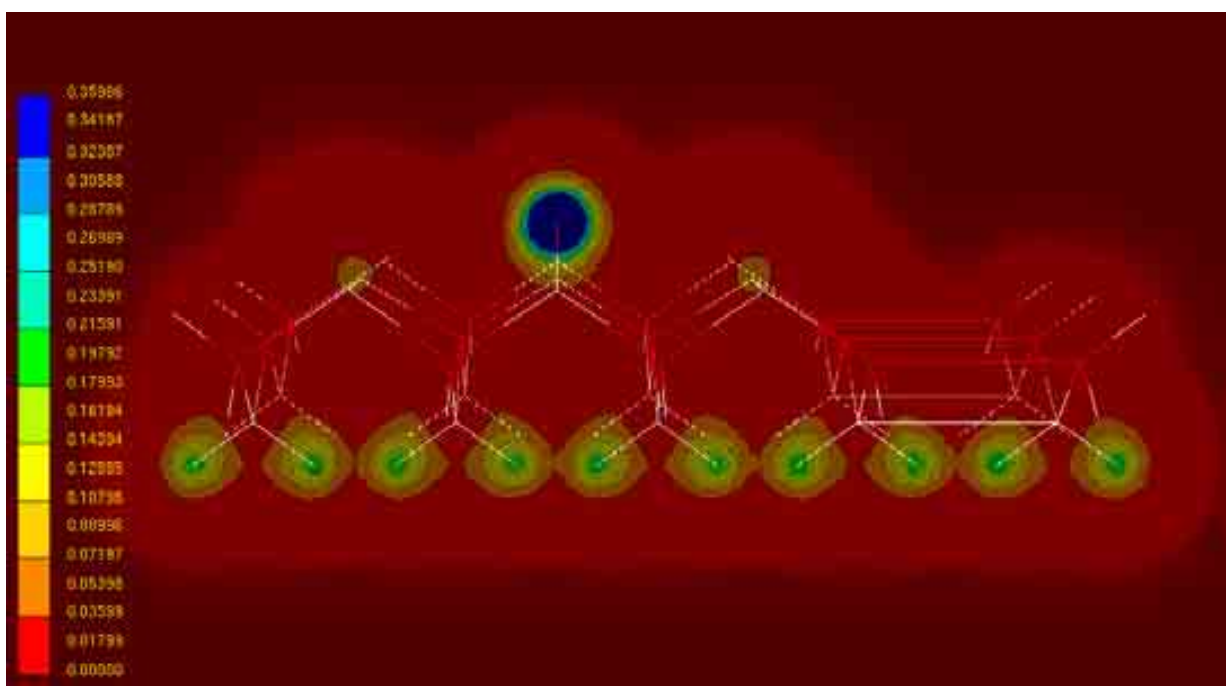

Figure 19. Electron charge distribution through a plane for the $Ga_{19}As_{15}H_{39}$ + O bridge site.

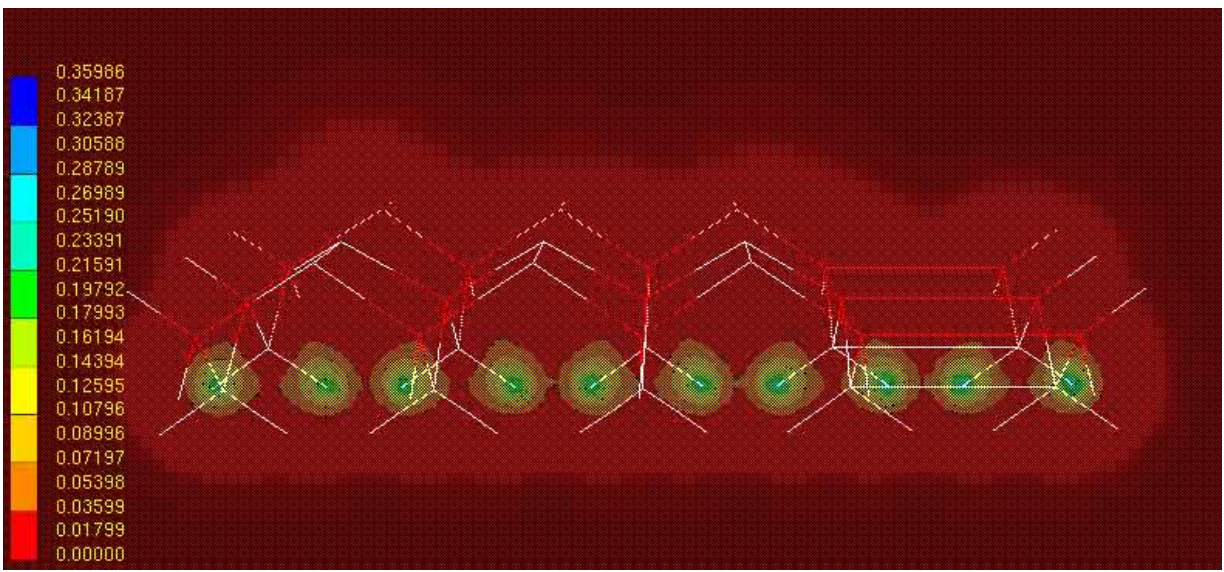

Figure 20. Electron charge distribution through a plane for the $Ga_{19}As_{15}H_{39}$ bare cluster.

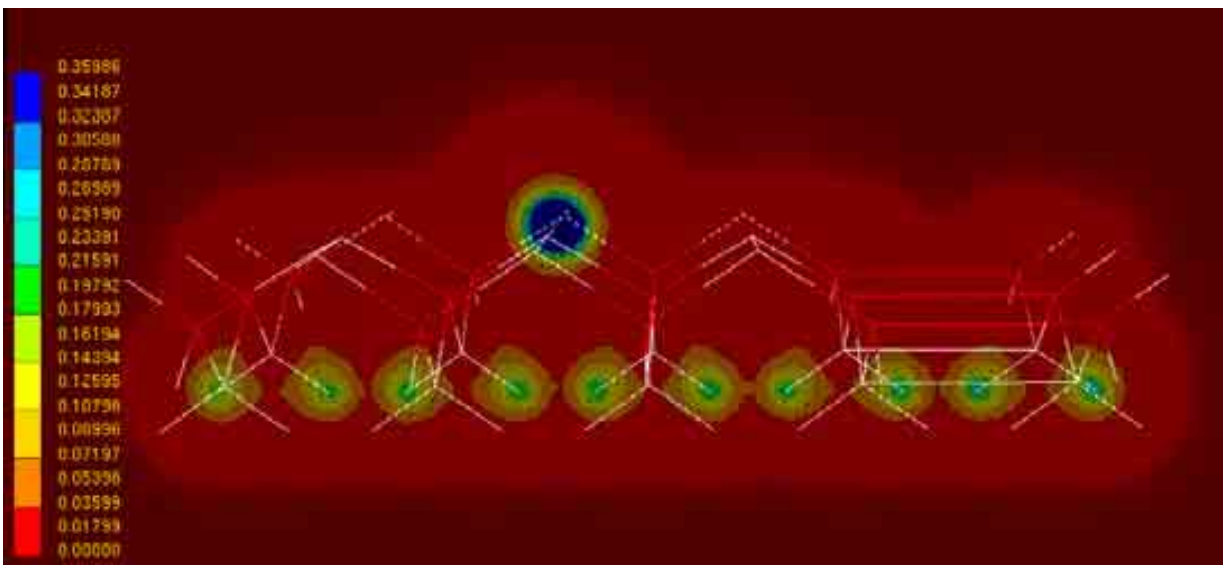

Figure 21. Electron charge distribution through a plane for the $Ga_{19}As_{15}H_{39}$ + O hollow site.